# FLUKA simulations of the target thickness dependence of Cu-Kβ/Kα intensity ratios

Aneta Maria Gójska[1], Karol Kozioł[1], Ewelina Agnieszka Miśta-Jakubowska[1], Adam Wasilewski[1], Krystian Trela[1]

[1]National Centre for Nuclear Research, ul. A. Soltana 7, 05-480 Otwock, Poland

**Abstract**:

The numerical Monte-Carlo simulations of Cu-Kα and Cu-Kβ fluorescence lines induced by monoenergetic $^{241}$Am radiation in copper have been presented. The simulations included modeling the Kβ/Kα intensity ratios for various thicknesses of copper. The results obtained using the FLUKA code were compared to available experimental and theoretical values. A clear relationship was observed between the simulated Kβ/Kα intensity ratios and the sample thickness: as the thickness increased, the Kβ/Kα ratio also increased until it reached saturation.

**Introduction**:

The X-ray spectroscopy plays a fundamental role in atomic research and is utilized in various fields, including atomic physics [1-17], material science, astrophysics [18-19], as well as in industry and physicochemical analyses in archaeometry [20-22]. Basic atomic parameters, such as intensity ratios [23-53], cross sections for X-ray production [12-17], fluorescence yields [11,15,16,31], or vacancy transfer probabilities [4-11, 32,47], play a crucial role in atomic physics. Experimental data often serve as comparative tests for theoretical models [1-3] and programs that simulate the interaction between radiation and matter [22], and vice versa. Among these basic atomic parameters, intensity ratios are particularly important due to their significance in the quantitative analysis of materials and their essential role in determining cross sections for excitation, ionization, and X-ray production. They are of great importance in atomic physics, for instance, in determining the probability of vacancy transfer (e.g., transferring a vacancy from the K shell to the L shell).

The study of Kβ/Kα intensity ratios can provide valuable information about the structure and chemical composition of materials. According to Daoudi et al. [54] and Hamidani [55], this field has witnessed over a thousand measurements and has resulted in a substantial number of theoretical and experimental publications over the past fifty years. Therefore, it has a wide range of applications in various scientific fields and industries. Here are a few examples that motivate further research on Kβ/Kα intensity ratios:

1. In archaeometry [20-22]: analyzing the Kβ/Kα intensity ratio can be applied to the chemical analysis of ancient artifacts such as coins and jewelry, allowing for the determination of their geographical and chronological origins. These studies have recently gained significant interest as archaeometry is an interdisciplinary field that combines knowledge from different disciplines such as physics, chemistry, statistics, and archaeology. This multidisciplinary approach enables better research outcomes, leading to a more accurate understanding of the history and culture of our ancestors. The Kβ/Kα intensity ratio can also be used to study surface enrichment of silver [56].

2. In astrophysics [18,19]: Kβ/Kα intensity ratios have particular applications in studying supernova remnants (SNRs). Iron spectra in various charge states (Fe XVII through Fe XXV) are most commonly observed. The obtained Kβ/Kα ratios allow for the determination of electron density, plasma temperature, and degree of ionization. The Fe Kβ/Kα intensity ratios are sensitive to the Fe ionic charge because the fluorescence efficiency of these lines depends on the number of bound electrons in the 2p and 3p shells.

3. In materials science: investigating the Kβ/Kα intensity ratio can provide information about the structure and chemical composition of materials, which is essential in designing and producing new materials with desired properties. Furthermore, an important application of this technique is quality control and elemental identification in various materials such as mineral resources, plastics, and food products. This allows for ensuring product safety and quality while avoiding fraud. The examination of the Kβ/Kα intensity ratio can be applied in industries for evaluating the quality of electroplated coatings or protective coatings used on metallic machine and equipment components.

In our previous work [22] we observed significant discrepancies between a set of experimental data. The experimental results fell between the FLUKA simulation [57, 58] results obtained for very thin (1 μm) and very thick (2 mm) samples, indicating that the differences in the experimental results for pure copper were likely due to variations in sample thickness used in the experiments. In this study we conducted Monte-Carlo FLUKA simulations for varying copper thicknesses. Theoretical Multi-Configuration Dirac-Hartree-Fock (MCDHF) calculations were also performed for various electron configurations. The obtained results were compared with available experimental and theoretical values.

**Experiment simulations**

The simulations were conducted using the FLUKA 2011 code version 2c.8, which was installed on a computer cluster at the Świerk Computing Center [59]. The FLUKA code utilizes the Evaluated Photon Data Library (EPDL97) [60], which contains detailed tables of photon interaction data encompassing photoionization, photoexcitation, coherent and incoherent scattering, as well as pair and triplet production cross sections.

The experimental setup reproduced in the calculations, consisting of monoenergetic $^{241}$Am (59.9 keV) radiation, and irradiated a sample groups with a diameter of 1 cm and a thickness of 0.5 μm, 1 μm, 10 μm, 50 μm, 200 μm, 500 μm, 1 mm, 2 mm, 4 mm, is shown in Figure 1.

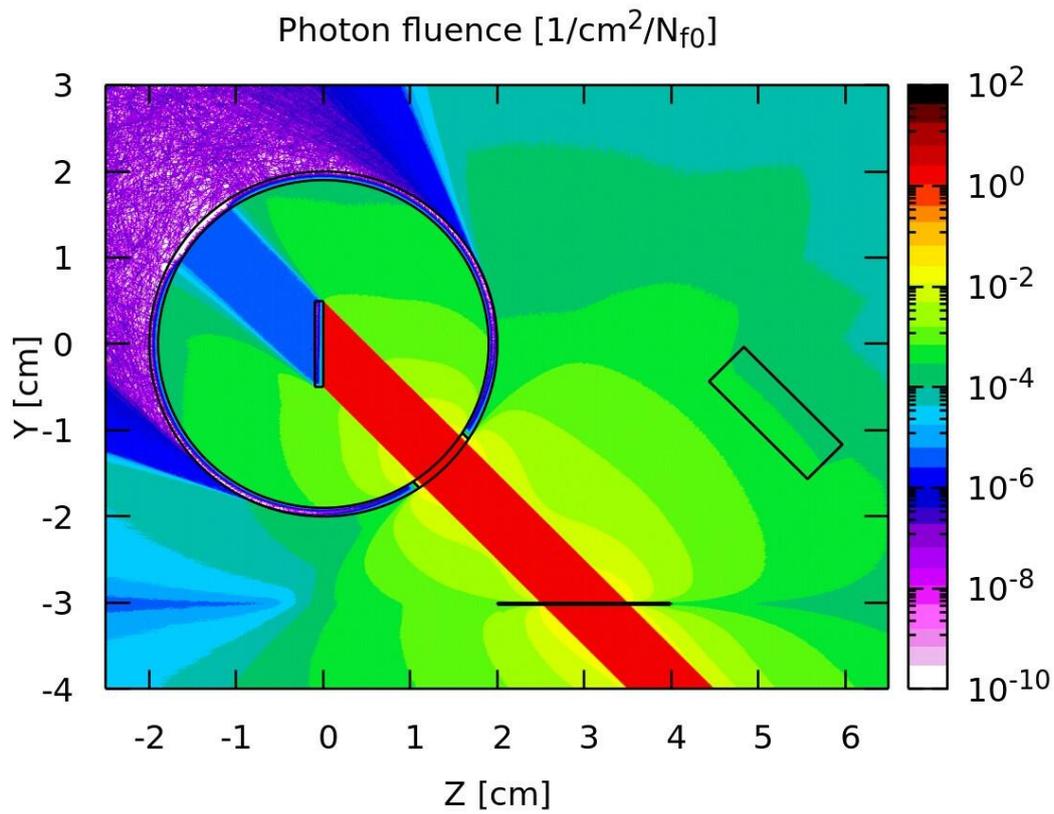

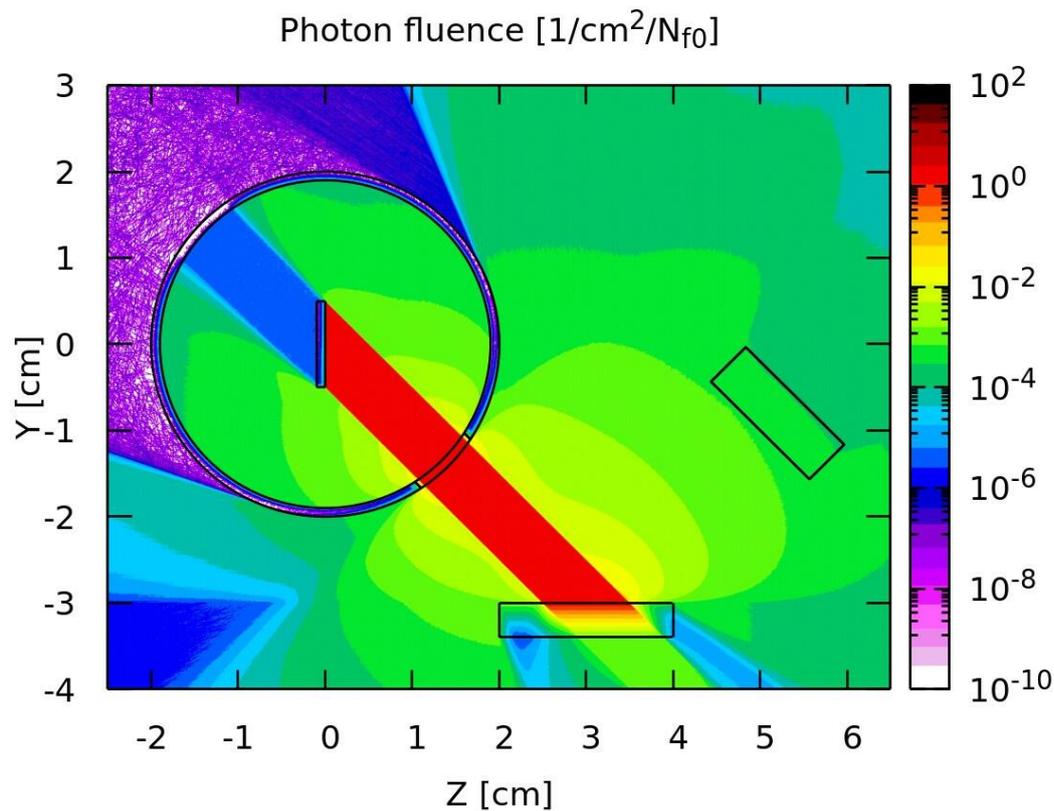

Fig. 1. Experimental setup and photon fluence reproduced in the calculations for the thin (up) and thick (bottom) samples.

As depicted in Figure 1, when dealing with thin samples, a portion of radiation passes through the sample while another portion is reflected back from the sample. The former is referred to as the forward output flux, while the latter is known as the backward output flux (typically directed towards the detector). However, when dealing with thick samples, there is no forward output flux since all radiation passing through the sample gets absorbed or re-emitted in a backward direction. Figure 2 showcases the computed K X-ray spectra of Cu, which were observed on a flat surface of an irradiated sample. As one can see from Figure 2, in the case of thin (10 µm) sample the forward and the backward output fluxes are comparable in size, but in the case of thick (1 mm) sample the forward flux is significant smaller then the backward output flux.

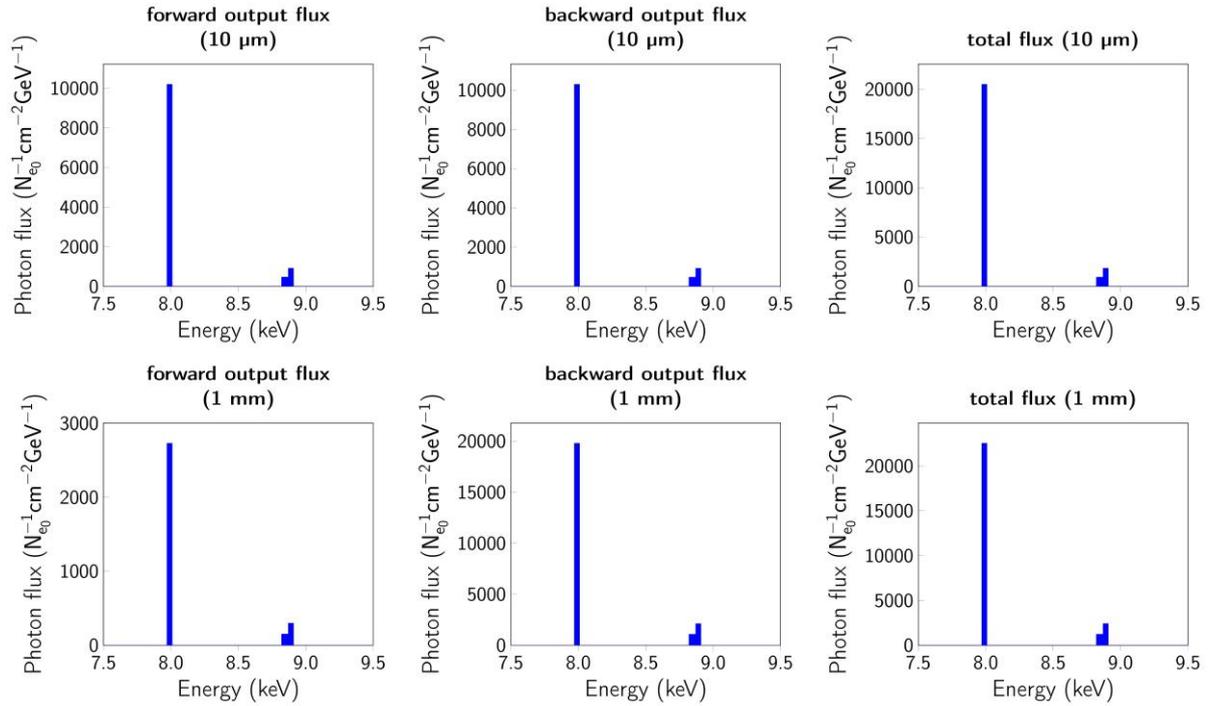

Fig 2. Example simulated X-ray spectra of Cu with varying thicknesses.

**Results and discussion**

Figure 3 provides a comprehensive overview of Cu-Kβ/Kα intensity ratio measurements conducted between 1976 and 2020 including the corresponding theoretical values. The measurements were performed using various experimental methods and under different conditions. Most of the studies were conducted by irradiating the sample with a radioactive source of $^{241}$Am (energy 59.9 keV) [16, 23-37, 48, 50-52], of $^{109}$Cd (22.69 keV) [30-41], and of $^{137}$Cs (32.86 keV) [42]. However, the results obtained for X-ray tube [38,43,45,46], K-electron capture [43], protons (1 MeV, 2 MeV) [44], synchrotron radiation (10 keV) [47] and high-energy electron beam [53] are also available. The Cu-Kβ/Kα values were obtained by use of Si(Li) [16,24,26,28-33, 35,37-41,43-45,47,48,50], Ge [23,25, 27,34,36,42,51,52] and high-resolution double-crystal X-ray spectrometer with a proportional counter Si(220) [46].

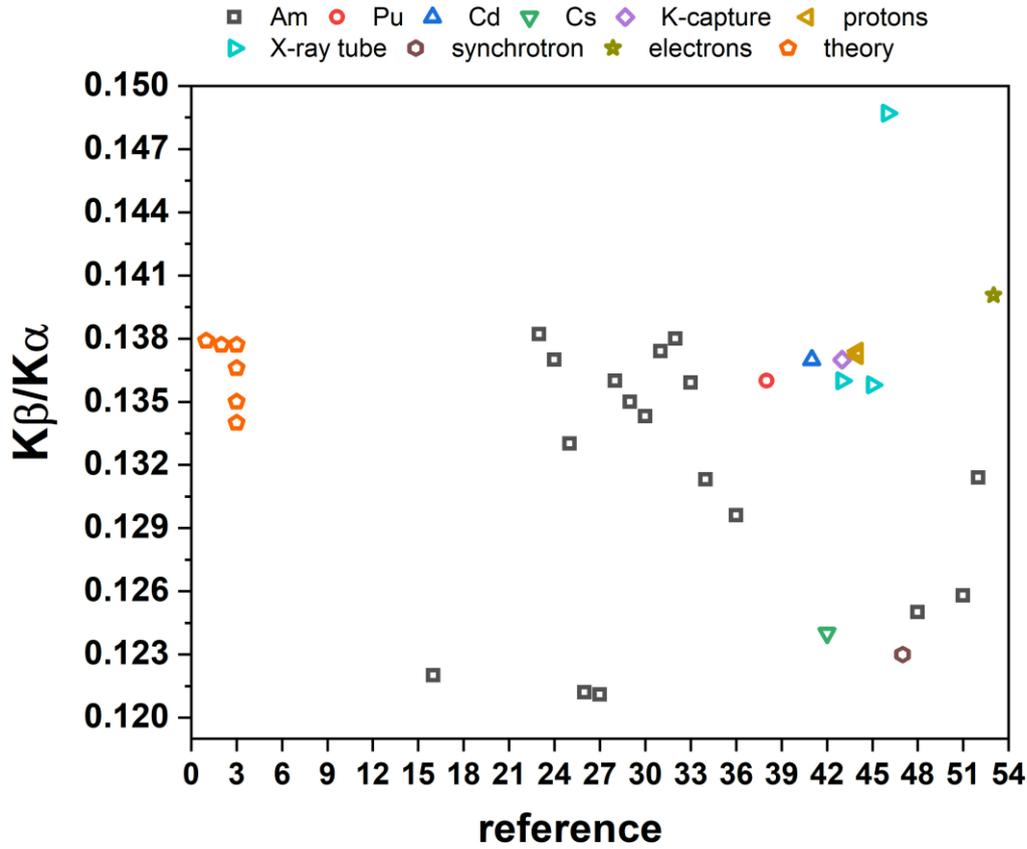

Fig. 3. Theoretical and experimental Cu-Kβ/Kα intensity ratios according to the references.

In our study, we utilized FLUKA simulations to obtain data, which are presented in Figure 4 alongside values from existing literature. Our MCDHF calculations for various valence electronic configuration and for OIE effect are also included (see Table 1). The experimental results fall within the range of FLUKA results for 0.5 μm and 50 μm, while the theoretical calculations fall within the range of 1 μm and 10 μm. This indicates that the Cu-Kβ/Kα intensity ratio is sensitive to sample thickness. Moreover, the intensity ratio of Cu-Kβ/Kα increases with increasing thickness and reaches saturation at the value of 0.2 mm.

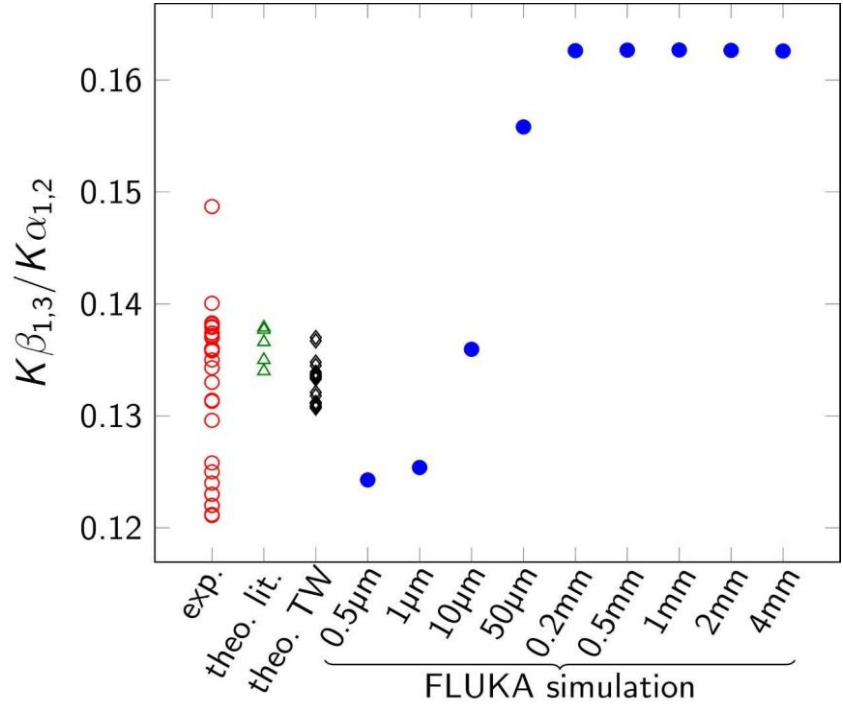

Fig. 4. The comparison between Cu-Kβ/Kα values obtained from the literature (exp. - experimental and theo. lit. - theoretical), theoretical calculations in this work (theo. TW), and those obtained using FLUKA for various sample thickness.

The MCDHF method have been applied to calculations of Kβ/Kα intensity ratios for Cu. The GRASP2018 code [61] has been used. The methodology of MCDHF calculations performed in the present study is similar to that published earlier in many papers (see, e.g., [62,63]). Our MCDHF calculations were performed for various valence electronic configuration (see Table 2), such as configurations of not ionized valence shells: $3d^{10}4s^1$, $3d^{10}4p^1$, $3d^9 4s^2$, configurations with partially ionized valence shells: $3d^{10}$, $3d^9 4s^1$, $3d^9 4p^1$, $3d^9$, $3d^8 4s^2$, and configuration with additional valence electron: $3d^{10}4s^2$. The calculations of transition rates were performed for Coulomb and Babushkin gauges. We also studied the effect of satellite contributions influence on lines ratio, so called the outer-shell ionization and excitation (OIE) effect [64,65,66]. Ionization and excitation processes following the K-shell ionization lead to a contribution of the configurations with additional ionization in the valence shells amongst the deexciting atoms and, in consequence, to a changing of $K\beta_{1,3}/K\alpha_{1,2}$ ratio. In order to evaluate the OIE effect on the $K\beta_{1,3}/K\alpha_{1,2}$ ratio the calculations of the total shake probabilities, i.e., shake-off and shake-up, are calculated by applying the sudden approximation model [67] and using wave functions calculated by GRASP2018. See [64,67] for more detailed description of the procedure of including OIE effect in the analyze of X-ray spectra.

As one can see from Table 1, from an atomic theory point of view the Kβ/Kα intensity ratio depends mostly on the number of the 3d electrons: the lower number of 3d electrons, the higher Kβ/Kα ratio. This finding is supported by the calculations of Polasik [3] performed for two valence configurations of Cu: $3d^{10}4s^1$ and $3d^9 4s^2$. The OIE effect leds to a mixture of

$3d^n4s^x$ and $3d^{n-1}4s^x$ configurations what results in the higher Kβ/Kα ratio comparing to the "pure" $3d^n4s^x$ configuration.

Table. 1. MCDHF calculations were performed for various valence electronic configuration

| Valence configuration | $K\beta_{1,3}/K\alpha_{1,2}$ (Babushkin) | $K\beta_{1,3}/K\alpha_{1,2}$ (Coulomb) |
|---|---|---|
| $3d^{10}4s^1$ | 0.1308 | 0.1311 |
| $3d^{10}4p^1$ | 0.1308 | 0.1312 |
| $3d^{10}$ | 0.1307 | 0.1311 |
| $3d^{10}4s^2$ | 0.1309 | 0.1312 |
| $3d^94s^1$ | 0.1334 | 0.1337 |
| $3d^94p^1$ | 0.1335 | 0.1339 |
| $3d^9$ | 0.1336 | 0.1339 |
| $3d^94s^2$ | 0.1333 | 0.1337 |
| $3d^84s^2$ | 0.1367 | 0.1370 |
| $3d^{10}4s^1$ + satellites (OIE) | 0.1318 | 0.1321 |
| $3d^94s^2$ + satellites (OIE) | 0.1345 | 0.1348 |

**Conclusions**:

The Cu-Kβ/Kα intensity ratios for copper with various thicknesses have been simulated using the FLUKA code. The results demonstrate that the Cu-Kβ/Kα intensity ratio is influenced by the thickness of the sample. In the range of 0.5–50 μm for copper sample thickness it is possible to link quantitatively the Kβ/Kα intensity ratio and layer thickness and thus use the Kβ/Kα ratio to examine the sample thickness. These findings have significant implications in fields such as archaeometry [4, 22], calibration of standards [68], and determination of depth profiles [69,70].

**References**:

[1] J.H Scofield, Exchange corrections of K x-ray emission rates. Phys. Rev. A 9 (1974) 1041–1049.
[2] K Jankowski, M. Polasik, On the calculation of Kβ/Kα X-ray intensity ratios. J. Phys. At. Mol. Opt. Phys. 22 (1989) 2369–2376.
[3] M. Polasik, Influence of changes in the valence electronic configuration on the Kβ-to-Kα X-ray intensity ratios of the 3d transition metals. Phys. Rev. A 58 (1998) 1840–1845.

[4] A.M Gójska, K. Kozioł, E.A. Miśta-Jakubowska, R. Diduszko, Determination of the Kβ/Kα intensity ratios of silver in Ag-Cu alloys. Nucl. Instrum. Methods Phys. Res. Sect. B Beam Interact. Mater. Atoms. 468 (2020) 65–70.

[5] B. Ertugral, U. Cevik, E. Tirasoglu; A. Kopya, M. Ertugrul, O. Dogan, Measurement of K to L shell vacancy transfer probabilities for the elements 52≤Z≤68. J. Quant. Spectrosc. Radiat. Transf. 78, (2003) 163–169.

[6] B. Ertuğral, H. Baltaş, A. Çelik, Y. Kobya, Vacancy Transfer Probabilities from K to L Shell for Low Atomic Number Elements at 5.96 keV. Acta Phys. Pol. A. 117 (2010) 900–903.

[7] S. Puri, D. Mehta, B. Chand, N. Singh, P. Trehan, Measurements of K to L shell vacancy transfer probability for the elements 37≤Z≤42. Nucl. Instrum. Methods Phys. Res. Sect. B Beam Interact. Mater. Atoms. 73 (1993) 443–446.

[8] E. Öz, Determination of ratios of emission probabilities of Auger electrons and K–L-shell radiative vacancy transfer probabilities for 17 elements from Mn to Mo at 59.5 keV. J. Quant. Spectrosc. Radiat. Transf. 97 (2006) 41–50.

[9] L.F.M Anand, S.B. Gudennavar, S.G. Bubbly, B.R. Kerur, Kβ to Kα X-ray intensity ratios and K to L shell vacancy transfer probabilities of Co, Ni, Cu, and Zn. J. Exp. Theor. Phys. 121 (2015) 961–965.

[10] S. Mirji, A. Bennal, Krishnananda, N. Badiger, M. Tiwari, G. Lodha, Determination of K-L vacancy transfer probabilities of some 3d elements using synchrotron radiation. Can. J. Phys. 93 (2015) 760–764.

[11] Ö. Söğüt, E. Büyükkasap, A. Küçükönder, *et al.* Measurement of vacancy transfer probability from K to L shell using K-shell fluorescence yields. Pramana - J Phys. 73 (2009) 711–718.

[12] R. Durak, Y Özdemır, Measurement of K-shell fluorescence cross-sections and yields of 14 elements in the atomic number range 25≤Z≤47 using photoionization. Radiat. Phys. Chem. 61 (2001) 19–25.

[13] M. Şahin, L. Demir, G. Budak, Measurement of K X-ray fluorescence cross-sections and yields for 5.96 keV photons. Appl. Radiat. Isot. 63 (2005) 141–145.

[14] E. Baydas, Y. Şahin, E. Büyükkasap, Measurement of *Kα* and *Kβ* X-ray fluorescence cross-sections and the Kβ/Kα intensity ratios for elements in the range 22⩽*Z*⩽29 by photons, J. Quant. Spectrosc. Rad. 77 (2003), p. 87

[15]. R. Yilmaz, H.Tunç, A. Özkartal, Measurementsof K-shell X-ray production cross-sections and fluorescence yields for some elements in the atomic number range 28≤Z≤40, Radiation Physics and Chemistry. 112 (2015) 83–87

[16] I. Han, M. Şahin, L. Demir, Y. Şahin, Measurement of K X-ray fluorescence cross-sections, fluorescence yields and intensity ratios for some elements in the atomic range 22≤Z≤68 Applied Radiation and Isotopes. 65 (2007) 669–675

[17] M.R. Kaçal, I. Han, F. Akman, Determination of K shell absorption jump factors and jump ratios of 3d transition metals by measuring K shell fluorescence parameters. Appl. Radiat. Isot. 95 (2015) 193–199

[18] Hiroya Yamaguchi, P. John, Hughes, Carles Badenes, Eduardo Bravo, R. Ivo, Seitenzahl, Héctor Martínez-Rodríguez, Sangwook Park, and Robert Petre, The Origin Of

The Iron-Rich Knot In Tycho's Supernova Remnant, The Astrophysical Journal 834 (2017) 124

[19] A. P. Rasmussen, E. Behar, S. M. Kahn, J.W. den Herder, and K. van der Heyden, The X-ray spectrum of the supernova remnant 1E 0102.2−7219, Astronomy&Astrophysics 365 (2001) L231-L236

[20]. A. Gójska, A.; Miśta-Jakubowska, E.; Banaś, D.; Kubala-Kukuś, A.; Stabrawa, I. Archaeological applications of spectroscopic measurements. Compatibility of analytical methods in comparative measurements of historical Polish coins. Measurement. 135 (2019) 869–874.

[21] E. Miśta-Jakubowska, R. Czech Błośska, W. Duczko, A. Gójska, P. Kalbarczyk, G. Żabiński, K. Trela, Archaeometric studies on early medieval silver jewellery from Central and Eastern Europe. Archaeol. Anthropol. Sci. 11 (2019) 6705–6723.

[22] A. Gójska, K. Kozioł, A. Wasilewski, E. Miśta-Jakubowska, P. Mazerewicz, J. Szymanowski, FLUKA Simulations of Kβ/Kα Intensity Ratios of Copper in Ag–Cu Alloys. Materials. 14 (2021) :4462.

[23] E. Casnati, A. Tartari, C. Baraldi; G. Napoli, Measurement of the Kβ/Kα yield ratios of Cu, Mo and Cd stimulated by 59.54 keV photons. J. Phys. B At. Mol. Phys. 18 (1985) 2843–2849.

[24] U. Çevik, I. Değirmencioğlu, B. Ertuğral, G. Apaydın, H. Baltaş, Chemical effects on the Kβ/Kα X-ray intensity ratios of Mn, Ni and Cu complexes. Eur. Phys. J. D , 36 (2005) 29–32.

[25] L.F.S. Coelho, M.B. Gaspar, J. Eichler, Kβ-to-Kα x-ray intensity ratios after ionization by γ rays. Phys. Rev. A. 40 (1989) 4093–4096.

[26] R. Yılmaz, Kβ/Kα X-ray intensity ratios for some elements in the atomic number range 28≤Z≤39 at 16.896 keV. J. Radiat. Res. Appl. Sci. 10 (2017) 172–177.

[27] A. Küçükönder, Y. Sąhín, Büyükkasap, E. Dependence of the Kβ/Kα intensity ratio on the oxidation state. J. Radioanal. Nucl. Chem. Artic. 170 (1993) 125–132.

[28] S. Raj, B.B. Dhal, H.C. Padhi, M. Polasik, Influence of solid-state effects on the Kβ-to-Kα x-ray intensity ratios of Ni and Cu in various silicide compounds. Phys. Rev. B. 58 (1998) 9025–9029.

[29] M. Ertuğrul, Ö. Söğüt, Ö. Şimşek, E. Büyükkasap, Measurement of Kβ/Kα intensity ratios for elements in the range 22≤Z≤69 at 59.5 keV. J. Phys. B At. Mol. Opt. Phys. 34 (2001) 909–914.

[30] S. Raj, H.C. Padhi, P. Palit, D.K. Basa, M. Polasik, F. Pawłowski, Relative K x-ray intensity studies of the valence electronic structure of 3d transition metals. Phys. Rev. B. 65 (2002) 193105.

[31] Ö. Söğüt, E. Baydaş, E. Büyükkasap, Y. Şahin, A. Küçükönder, Chemical effects on L shell fluorescence yields of Ba, La and Ce compounds. J. Radioanal. Nucl. Chem. 251 (2002) 119–122.

[32] E. Öz, Determination of ratios of emission probabilities of Auger electrons and K–L-shell radiative vacancy transfer probabilities for 17 elements from Mn to Mo at 59.5 keV. J. Quant. Spectrosc. Radiat. Transf. 97 (2006) 41–50.

[33] B. Ertuğral, G. Apaydın, U. Çevik, M. Ertuğrul, A.I. Kobya, Kβ/Kα X-ray intensity ratios for elements in the range 16≤Z≤92 excited by 5.9, 59.5 and 123.6 keV photons. Radiat. Phys. Chem. 76 (2007) 15–22.
[34] N.K. Aylikci, E. Tiraşoğlu, G. Apaydın, E. Cengiz, V. Aylikci, Bakkaloğlu, Ö.F. Influence of alloying effect on X-ray fluorescence parameters of Co and Cu in CoCuAg alloy films. Chem. Phys. Lett. 475 (2009) 135–140.
[35] T. Akkuş, Y. Şahin, D. Yılmaz, F.N. Tuzluca, The K-beta/K-alpha intensity ratios of some elements at different azimuthal scattering angles at 59.54 keV. Can. J. Phys. 95 (2017) 220–224.
[36] M. Dogan, M. Olgar, E. Cengiz, E. Tıraşoglu, Alloying effect on K shell X-ray fluorescence cross-sections and intensity ratios of Cu and Sn in $Cu_1Sn_{1-x}$ alloys using the 59.5 keV gamma rays. Radiat. Phys. Chem. 126 (2016) 111–115.
[37] M. Uğurlu, L. Demir, Relative K X-ray intensity ratios of the first and second transition elements in the magnetic field. J. Mol. Struct. 1203 (2020) 127458.
[38] N. Venkateswara Rao, S. Bhuloka Reddy, G. Satyanarayana, D. Sastry, Kβ/Kα X-ray intensity ratios in elements with 20≤Z≤50. Physica B+C. 138 (1986) 215–218.
[39] S. Porikli, Y. Kurucu, Effect of an External Magnetic Field on the K α and K β x-Ray Emission Lines of the 3d Transition Metals. Instrum. Sci. Technol. 36 (2008) 341–354.
[40] M.R. Kaçal, İ. Han, F. Akman, Determination of K shell absorption jump factors and jump ratios of 3d transition metals by measuring K shell fluorescence parameters. Appl. Radiat. Isot. , 95 (2015) 193–199.
[41] I. Han, L. Demir, Effect of annealing treatment on Kβ-to-Kα x-ray intensity ratios of 3D transition-metal alloys. Phys. Rev. A. 81 (2010) 062514
[42] L.F.M. Anand, S.B. Gudennavar, S.G. Bubbly, B.R. Kerur, K-Shell X-ray Fluorescence Parameters of a Few Low Z Elements. J. Exp. Theor. Phys. 126 (2018) 1–7.
[43] G. Paić, V. Pečar, Study of anomalies in Kβ/Kα ratios observed following K-electron capture. Phys. Rev. A. 14 (1976) 2190–2192.
[44] A. Perujo, J.A. Maxwell, W.J. Teesdale, J.L. Campbell, Deviation of the Kβ/Kα intensity ratio from theory observed in proton-induced X-ray spectra in the 22 ≤ Z ≤ 32 region. J. Phys. B At. Mol. Phys. 20 (1987) 4973–4982
[45] M.M. Bé, M.C. Lépy, J. Plagnard, B. Duchemin, Measurement of relative X-ray intensity ratios for elements in the 22≤Z≤29 Region. Appl. Radiat. Isot. 49 (1998) 1367–1372.
[46] Y. Ito, T. Tochio, M. Yamashita, S. Fukushima, T. Shoji, K. Słabkowska, Ł. Syrocki, M. Polasik, J. Padežnik Gomilsek, J. P. Marques, J. Sampaio, M. Guerra, J. Machado, J. P. Santos, A. Hamidani, A. Kahoul, P. Indelicato and F. Parente, Intensity Ratio of Kβ∕Kα in Selected Elements from Mg to Cu, and the Chemical Effects of Cr $Kα_{1,2}$ Diagram Lines and Cr Kβ∕Kα Intensity Ratio in Cr Compounds, Int J Mol Sci. 24(6) (2023) 5570.
[47] S. Mirji, A. Bennal, Krishnananda, N. Badiger, M. Tiwari, G. Lodha, Determination of K-L vacancy transfer probabilities of some 3d elements using synchrotron radiation. Can. J. Phys. 93 (2015) 760–764.
[48] B.G. Durdu, U. Alver, A. Küçükönder, Ö. Söğüt and M. Kavgaci, Investigation on Zinc Selenide and Copper Selenide Thin Films Produced by Chemical Bath Deposition, Acta Physica Polonica A Vol. 124 (2013) 41-45

[49] G.Apaydin, V.Aylikçi, Z.Biyiklioğlu, E.Tiraşoğlu, H.Kantekin, Influence of Chemical Effect on the Kβ/Kα Intensity Ratios and Kβ Energy Shift of Co, Ni, Cu, and Zn Complexes. Chinese Journal of Chemical Physics, 21(6) (2008) 591-595

[50] L. Demir, Y. Şahin, Measurement of K x-ray fluorescence parameters in elements with 24≤Z≤65 in an external magnetic field, Radiation Physics and Chemistry 85 (2013) 64-69

[51] V. Aylikci, A. Kahoul, Aylikci, N. Kup, E.Tiraşoğlu, İ. H. Karahan, Empirical, Semi-Empirical and Experimental Determination of K X-Ray Fluorescence Parameters of Some Elements in the Atomic Range 21 ≤ Z ≤ 30. Spectroscopy Letters. 48(5) (2015) 331–342. doi:10.1080/00387010.2014.881381

[52] A. Küçükönder, Y. Şahin, E. Büyükkasap, A. Kopya, Chemical effects on Kβ/Kα X-ray intensity ratios in coordination compounds of some 3d elements,
J. Phys. B At. Mol. Opt.Phys. 26 (1993b) 101–105

[53]. D Berenyi, G Hock, S Ricz, B Schlenk and A Valek, Kα/Kβ x-ray intensity ratios and K-shell ionisation cross sections for bombardment by electrons of 300-600 keV, J. Phys. B: Atom. Molec. Phys. 11 (1978) 709-713

[54] S. Daoudi, A. Kahoul, N. Kup Aylikci, J. Sampaio, J. Marques, V. Aylikci, Y. Sahnoune, Y. Kasri, B. Deghfel, Review of experimental photon-induced Kβ/Kα intensity ratios. At. Data Nucl. Data Tables. 132 (2020) 101308.

[55] A. Hamidani, S. Daoudi, A. Kahoul , J.M. Sampaio, J.P. Marques, F. Parente,
S. Croft, A. Favalli, N. Kup Aylikci, V. Aylikci, Y. Kasri, K. Meddouh, Updated database, semi-empirical and theoretical calculation of Kβ/Kα intensity ratios for elements ranging from $^{11}$Na to $^{96}$Cm, Atomic Data and Nuclear Data Tables. 149 (2023) 101549

[56] A. Gójska, E. Miśta-Jakubowska, K. Kozioł, A. Wasilewski, R. Diduszko, The K-X-ray intensity ratios as a tool of examination and thickness measurements of coating layers. DOI: 10.48550/arXiv.2306.14526

[57] T. Böhlen, F. Cerutti, M. Chin, A. Fassò, A. Ferrari, P. Ortega, A. Mairani, P. Sala, G. Smirnov, V. Vlachoudis, The FLUKA Code: Developments and Challenges for High Energy and Medical Applications. Nucl. Data Sheets. 120 (2014) 211–214.

[58] A. Ferrari, p. Sala, A. Fasso, J. Ranft, FLUKA: A Multi-Particle Transport Code; Technical Report October; Stanford Linear Accelerator Center (SLAC): Menlo Park, CA, USA, 2005.

[59] Świerk Computing Centre. Infrastructure and Services for Power Industry. Available online: http://www.cis.gov.pl/ (accessed on 9 August 2021).

[60] D. Cullen, J. Hubbell, L. Kissel, EPDL97: The Evaluated Photo Data Library '97 Version; Technical report; Lawrence Livermore National Laboratory (LLNL): Livermore, CA, USA, (1997).

[61] C. Froese Fischer, G. Gaigalas, P. Jönsson, and J. Bieroń, Comput. Phys. Commun. 237 (2019). 184

[62] P. Jönsson, M. Godefroid, G. Gaigalas, J. Ekman, J. Grumer, W. Li, J. Li, T. Brage, I. P. Grant, J. Bieroń, and C. F. Fischer, Atoms 11 (2022).7

[63] I. P. Grant, Relativistic Quantum Theory of Atoms and Molecules (Springer, New York, NY. (2007).

[64] K. Kozioł, J. Quant. Spectrosc. Radiat. Transf. 149 (2014) 138 .

[65] M. Polasik, K. Słabkowska, J. Rzadkiewicz, K. Kozioł, J. Starosta, E. Wiatrowska-Kozioł, J.-C. Dousse, and J. Hoszowska, Phys. Rev. Lett. 107 (2011) 073001 .

[66] K. Kozioł and J. Rzadkiewicz, Phys. Rev. A. 96 (2017) 031402 .

[67]. T. A. Carlson, C. W. Nestor, T. C. Tucker, and F. B. Malik, Phys. Rev. 169 (1968) 27 .

[68] X. She, Z. Zhu, J. Gao, R. Qian, C. Sheng, R. Shen, S. Zhuo, Application of Kβ/Kα in selecting calibration standards for X-ray fluorescence analysis, X-Ray Spectrometry. 48 (2019) 664-673.

[69]. M. Ahlberg, Simple depth profile determination by proton-induced x-ray emission. Nucl. Instrum. Methods. 131 (1975) 381–384.

[70] T. Trojek, T. Čechák, and L. Musílek, Kα/Kβ Ratios of Fluorescence X-rays as an Information Source on the Depth Distribution of Iron in a Low Z Matrix, Analytical Sciences 24 (2008), 851-854